\documentclass[aps,prb,twocolumn]{revtex4-1}
\usepackage{amsmath}
\usepackage{amssymb}
\usepackage{amsfonts}
\usepackage[justification=raggedright]{caption}
\usepackage{subcaption}
\usepackage{graphicx}
\usepackage{epsfig}
\usepackage{hyperref}
\usepackage{epstopdf}
\usepackage{gensymb}
\usepackage{color }
\usepackage{color, soul }
\usepackage{xcolor}
\hypersetup{
colorlinks=true,
urlcolor=blue,
citecolor=blue,
linkcolor=blue,}

\begin{document}
\title{Observation of planar Hall effect in the magnetic Weyl semimetal Co$_{3}$Sn$_{2}$S$_{2}$} 

\author{Shama, R. K. Gopal, and Yogesh Singh}
\affiliation{Department of Physical Sciences,
Indian Institute of Science Education and Research Mohali,
Sector 81, S. A. S. Nagar, Manauli, PO: 140306, India}

\date{\today}

\begin{abstract}
We report detailed magneto-transport measurements on single crystals of the magnetic Weyl semi-metal Co$_{3}$Sn$_{2}$S$_{2}$.  Recently a large anomalous Hall effect and chiral anomaly have been observed in this material which have been suggested to be related to the large Berry curvature between the Weyl points (Liu et al., Nature Physics (2018).). Another effect expected to result from the topological band structure of magnetic Weyl materials is the planar Hall effect (PHE). In this work we report observation of this intrinsic effect in single crystals of Co$_{3}$Sn$_{2}$S$_{2}$.  Crucially, the PHE is observed for temperature $T \leq 74$~K which is much smaller than the ferromagnetic ordering temperature $T_c = 175$~K\@.  Together with the large anomalous Hall conductivity, this further demonstrates the Topological character of Co$_3$Sn$_2$S$_2$.  

\end{abstract}

\maketitle

\section{Introduction}
During the last decade discovery of Topological insulators has opened a completely new direction in condensed matter physics. The study of these phases of matter have led to the discovery of other classes of topological phases distinct from the previous ones \cite{Hasan,Kane,Moore}. Topological insulators were discovered in the strongly spin orbit coupled materials characterized by a single two-dimensional Dirac dispersion on the surface. The Dirac dispersion in topological insulators lies in the bulk band gap and is topologically protected \cite{Hasan}. Very soon the gapless phases of topological semimetals were theoretically predicted and experimentally realized in many materials such as Cd$_3$As$_2$, Na$_3$Bi, TaAs and NbP to name a few  \cite{Armitage,Wan,Young,Burkova,Lv,Huang,Xu,Zhang,Shekhar,Borisenko}. In these semimetals the Dirac dispersion relation is three-dimensional, unlike two-dimensional in topological insulators.  In a Dirac Semimatal (DSM), conduction and valence band touch each other linearly at special momenta along symmetry axes in the Brillouin zone and are symmetry protected \cite{Armitage,Xu,Lv}. Breaking of symmetries such as inversion or time reversal leads to the Dirac point splitting into two Weyl points and such a state is known as Weyl Semimetal (WSM). Like DSM the low energy electronic excitations in the WSM follow the massless Dirac equation and are topologically protected against localization from disorder. The spin texture of the carriers in these semimetals, also known as spin momentum locking or spin chirality, prohibits them from backscattering from defects and disorder. WSMs show unusual signatures in transport properties such as very high mobility, low carrier density, giant linear magneto resistance (MR), anomalous hall effect and chiral anomaly etc.  An important signature of the chiral anomaly is the observation of negative longitudinal magneto-resistance (NLMR) in parallel magnetic and electric fields . The NLMR signature of the chiral anomaly has been observed in nonmagnetic DSMs such as TaAs, Na$_{3}$Bi, Cd$_{3}$As$_{2}$, and Bi$_{1-x}$Sb$_{x}$\cite{Hirschberger,Xiong,Zhang,Huang,H Li,Kim,Lv} .

Earlier version of the Weyl semimetals were mostly non-magnetic in nature. However, very recently a magnetic version of the Weyl state has been realized in some ferromagnetic semimetals and gapless anti-ferromagnetic semimetals \cite{Borisenko,Hirschberger,Sakai,Enke Liu,Qi Wang}.  Magnetism in these materials leads to time reversal symmetry being broken and Weyl points are naturally realized.  A large Berry curvature in these magnetic semimetals acts as an intrinsic magnetic field and results in a large anomalous Hall conductivity and Hall angle. Previously negative longitudnal magneto-resistance (NLMR) in collinear magnetic and electric fields has been regarded as a possible signature of the chiral anomaly expected in these WSMs \cite{H Li,Hirschberger,Zhang,Xiong,Kim}.  However many extrinsic mechanisms such as current jetting and crystal inhomogeneity can also result in a negative MR and a direct cause and effect link between the observation of a negative MR and the presence of the chiral anomaly is still debated \cite{Reis,Wiedmann,Breunig,Andreev,Assaf,Dai,Arnold}.  

Recent theoretical calculations on the WSM have predicted another transport signature of the chiral anomaly.  An unusual Hall effect is predicted and observed when a magnetic field is applied in the plane of the current and the voltage contacts \cite{Nandy,Burkov1,Taskin}. In this planar geometry the Lorentz force does not play any role.  Starting with a magnetic field parallel to the current $I$, as the magnetic field is rotated in the plane, the voltage shows an oscillatory behaviour with a minima and maxima at $45^\circ$ and $135^\circ$, respectively.  This effect is known as the planar Hall effect (PHE) as the Hall voltage is generated by the coplanar current and magnetic field.  A very large PHE has been observed in some DSM and WSM materials recently \cite{P. Li, Li, Singha,Kumar,Chen F,Wu,Yang J,Najmul}.  Such a large PHE has been proposed to result from a large Berry curvature between the Weyl nodes and is observed in both magnetic and nonmagnetic WSM \cite{Nandy,Burkov1}. 
A PHE is also observed in ferromagnetic thin films due to the anisotropic magnetization of the film \cite{Tang2003}.  However, the PHE if present due to ferromagnetism, tracks the magnetization and is observed for temperatures starting from the ferromagnetic critical temperature \cite{Tang2003}.

In this work we report observation of a large PHE in single crystals of the recently discovered magnetic Weyl semimetal Co$_{3}$Sn$_{2}$S$_{2}$ \cite{Enke Liu,Qi Wang}.  In contrast to other WSMs the Fermi level in Co$_3$Sn$_2$S$_2$ lies very close to the Weyl nodes ($60$~meV below the Weyl nodes). An important experimental observation in this magnetic Weyl semimetal is the large, intrinsic anomalous Hall conductivity.  This is in a stark contrast to the anomalous Hall conductivity observed in previous ferromagnetic metals, which includes contributions from extrinsic mechanisms such as skew scattering and side jump scattering \cite{Zhong,Naota}. In two dimensions a quantized value of anomalous Hall conductivity is observed in magnetic topological insulators \cite{Naota,Checkelsky,Chang,Kau}.  This signifies that the Weyl fermions are robust against scattering from phonons and defects in the crystals.  This robustness against scattering make these materials ideal for technological applications.  Thus, the discovery and detailed study of new magnetic WSMs is of fundamental and technological importance.

Unambiguous signatures of Topological character is difficult to obtain from transport measurements.  For example, the NLMR as a signature of the chiral anomaly, may be masked by other contributions, both intrinsic like orbital MR or extrinsic like current jetting effects.  On the other hand, recent experimental findings suggest that PHE can be observed even if signature of the chiral anomaly are masked by other contributions because the PHE does not depend on the position of the chemical potential from the Weyl nodes \cite{Singha,Yang J}. This implies that, irrespective of the position of the chemical potential and even in the presence of extrinsic effects, topological bands can still contribute to special transport effects such as PHE and therefore contributions of these topologically non-trivial bands can be probed by measurements other than NLMR.  Co$_3$Sn$_2$S$_2$ is an ideal candidate to test this hypothesis since the NLMR reported previously for this material is rather weak ($\approx 2~\%$) and is observed only at very low temperatures ($T = 0.3$~K) where other contributions are much smaller \cite{Enke Liu}.

We have therefore carried out detailed magneto-transport measurements on high quality single crystals of the newly proposed magnetic WSM Co$_3$Sn$_2$S$_2$ with the aim to check for the PHE as evidence of a magnetic WSM state.  We observe a large PHE in Co$_3$Sn$_2$S$_2$ which appears for temperature $T < ~70$~K, which is far below the ferromagnetic (FM) ordering temperature $T_c \approx 175$~K, strongly supporting the claim that the PHE is not coupled to the FM, but most likely originates from the chiral anomaly.

\section{Experiment Details}
\begin{figure}
\centering
\includegraphics[width=\linewidth]{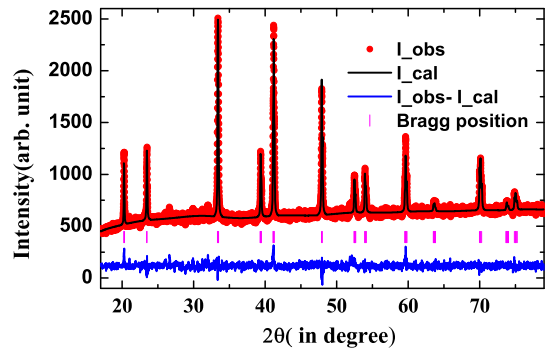}
\caption{The powder x-ray diffraction pattern and a Rietveld refinement of the same.. }
\label{fig-xrd}
\end{figure}

Single Crystals of Co$_{3}$Sn$_{2}$S$_{2}$ were grown by the Bridgman technique as in previous reports  \cite{Dedkov,Umetani,Holder,Kassem,Kassem2,Kassem3,Schnelle}. Stoichiometric amounts of Cobalt powder (Alpha Aesar 99.998~\%), Tin shots (Alfa Aesar 99.999~\%) and Sulfur pieces (Alfa Aesar 99.999~\%) were mixed in a mortal pestle, pelletized, and sealed in a quartz tube under vacuum. The sealed quartz tube was placed in a box furnace and heated to $1050~^{\circ}$C in $5$~hr, kept at this temperature for $6$~hr and then slowly cooled down to $800~^{\circ}$C in $72$~hr. Large shiny crystals are obtained by cleaving the as grown boule. The crystal structure was confirmed by powder X-ray diffraction (PXRD) on crushed single crystals, using a Rikagu diffractometer(Cu).  The PXRD data was refined using Full prof refinement technique. The refined lattice parameters are $a = b = 5.3666$~\AA , $c = 13.165$~\AA, which are close to values previously reported \cite{Enke Liu}.  The PXRD data and the refinement are shown in Fig.~\ref{fig-xrd}.  The stoichiometry of the crystals was confirmed by energy-dispersive x-ray spectroscopy (EDS) using a scanning electron microscope (SEM).  Transport measurements were performed in a Physical Property Measurement System (PPMS Quantum Design).  A four probe electrical contact configuration was adopted for the longitudinal resistivity and MR measurements.  For Hall and planar Hall effect measurements, a five probe configuration is used which nulls the contribution from the longitudinal MR. 

\begin{figure*}
\centering
\includegraphics[width=0.85\linewidth]{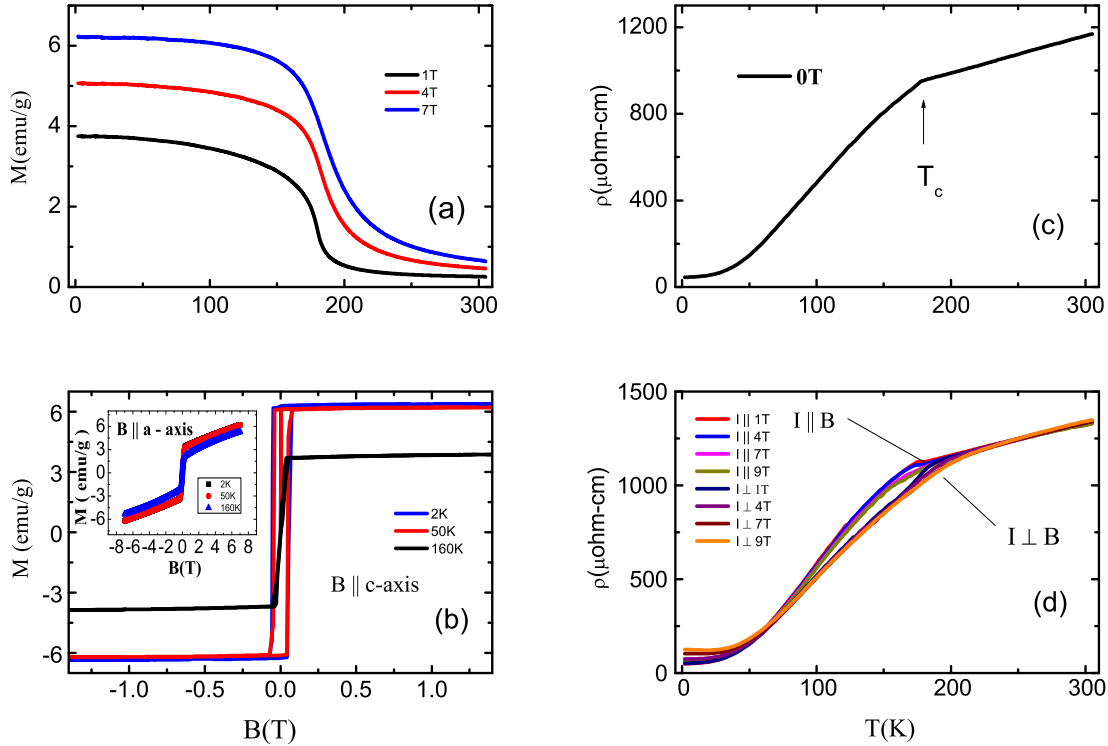}
\caption{(a) Temperature ($T$) dependent magnetization ($M$) data at different magnetic fields ($B$) (b) $M$ vs $B$ at various temperatures for $B \parallel c$-axis and (inset) $M$ vs $B$ for $B \parallel a$-axis (c) Resistivity $\rho$ vs $T$ data at $0$~T showing a kink at ferromagnetic transition temperature (d)  The $\rho(T)$ for various $B$ with current $I \parallel B$ and $I \perp B$. }
\label{fig:mt-rt}
\end{figure*}

\begin{figure*}
\centering
\includegraphics[width=\linewidth]{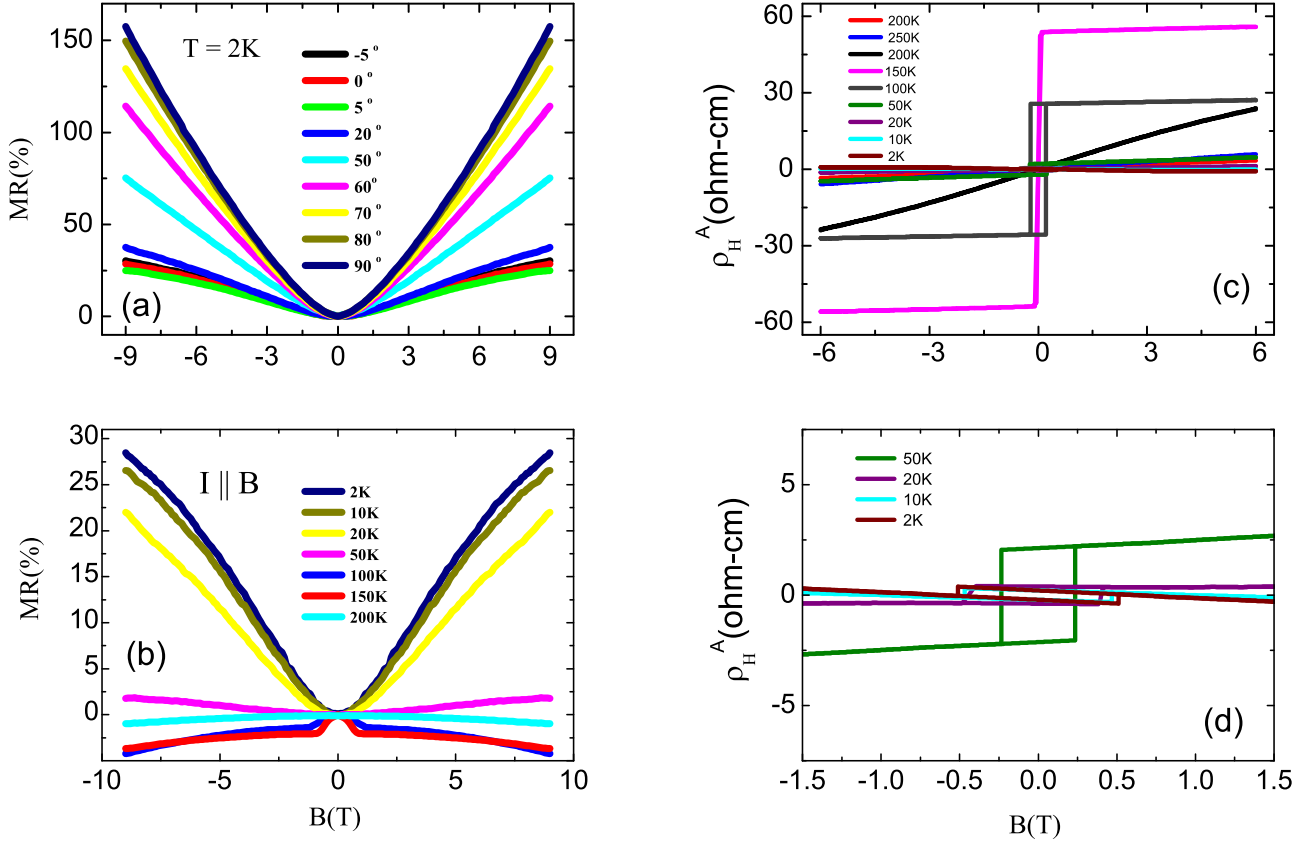}
\caption{(a) MR data at various angles between current $I$ and magnetic field $B$ (b) Longitudinal ($I \parallel B$) MR at various temperatures (c) Anomalous Hall resistivity $\rho^A_H$ vs $B$ at various Temperatures (d)  An expanded plot of $\rho^A_H$ vs $B$ to highlight the behaviour at low fields and temperatures.}
\label{fig:mr-hall}
\end{figure*}

\begin{figure*}
	\centering
	\includegraphics[width=0.85\linewidth]{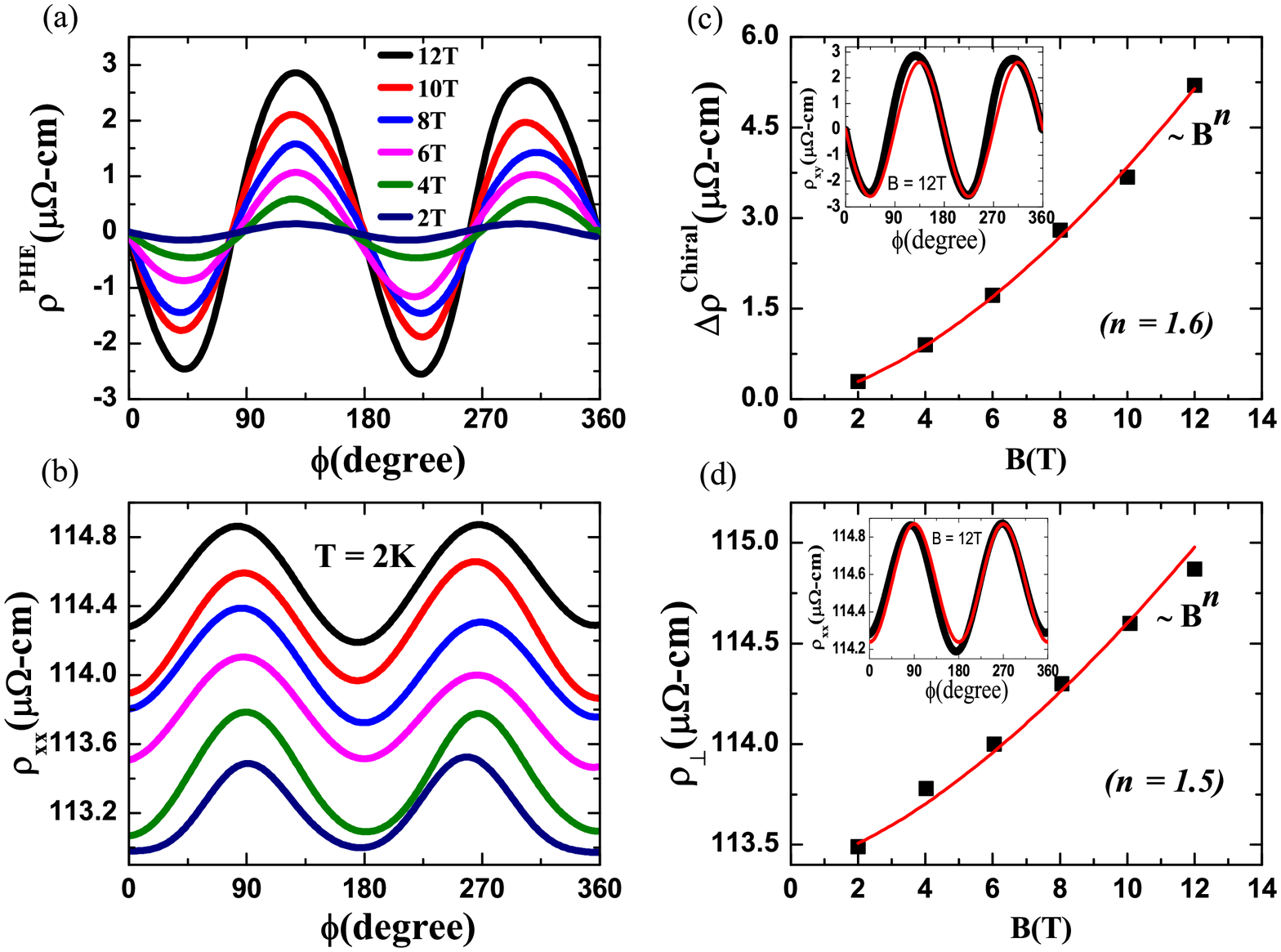}
	\caption{The (a) planar Hall resistivity $\rho^{PHE}$ and (b) the longitudinal resistivity $\rho_{xx}$ vs angle $\phi$ measured at $T = 2$~K in various magnetic fields. (c) Extracted chiral contribution $\triangle$$\rho_{chiral}$ vs magnetic field. (d) Extracted $\rho_{\perp}$ vs magnetic field. The solid curves through the data in (c) and (d) are fits to a power law field dependence $B^n$.  Insets in (c) and (d) show the fitting of $\rho^{PHE}$ and $\rho_{xx}$ data at $12$~T to the Eqns.~\ref{Eqn1} and \ref{Eqn2}, respectively.}
	\label{fig:planar}
\end{figure*}

\section{Results}
We first verify that our Co$_{3}$Sn$_{2}$S$_{2}$ single crystals show the expected magnetic and magneto-transport properties as have been reported previously \cite{Enke Liu}. Figure~\ref{fig:mt-rt} shows the basic magnetic and transport properties on Co$_3$Sn$_2$S$_2$ crystals.  From the temperature dependent magnetization data shown in Fig.~\ref{fig:mt-rt}~(a), it is clear that our single crystals show a ferromagnetic transition with a $T_c \approx 175$~K\@.  The field dependent magnetization data at various temperatures reveals a strong magnetic anisotropy between the in plane and out of plane field orientations as shown in Fig.~\ref{fig:mt-rt}~(b).  While the out of plane magnetization ($B \parallel c$) saturates in a small field $B = 0.1$~T, the in plane magnetization ($B \parallel ab$) does not saturate even up to $B = 7$~T (see inset in Fig.~\ref{fig:mt-rt}~(b)). This confirms that the magnetic ordering in this compound is dominantly out of plane, consistent with previous studies on this compound. 

Figure~\ref{fig:mt-rt}~(c) and (d) show the temperature dependent resistivity $\rho(T)$ data at various magnetic fields.  The $\rho(T)$ follows a typical metallic character with a value at $2$~K of $\rho(2K) = 45 \times 10^{-6}~\mu\Omega$~cm and at $300$~K a value of $\rho (300K) =  1168 \times 10^{-6}~\mu\Omega$~cm.  The residual resistivity ratio RRR $= \rho(300~{\rm K})/\rho(2~{\rm K}) = 26$ is significantly higher than the value of $8.8$ reported in ref \cite{Enke Liu} and indicates that our single crystals are of high quality.  The ferromagnetic transition at $T_c = 175$~K shows up as a kink in the $\rho(T)$ data measured in zero magnetic field as marked with an arow in Fig.~\ref{fig:mt-rt}~(c).  On the application of a magnetic field, the kink moves up in temperature and broadens as seen in the $\rho(T)$ data measured in magnetic fields up to $9$~T as shown in Figure~\ref{fig:mt-rt}~(d).

We next present MR and Hall data in the Fig.~\ref{fig:mr-hall}.  The MR data shown in Fig.~\ref{fig:mr-hall}~(a) is measured at $T = 2$~K with an excitation current $I = 1$~mA in the $ab$-plane of the crystal and the magnetic field at various out of plane angles to the direction of $I$.  For the angle $90^\circ$ ($B\parallel c$) configuration, the MR is positive and non-saturating, reaching a value of $160\%$ at a field of $9$~T as seen in Fig.~\ref{fig:mr-hall}~(a). This value is higher than the value previously reported \cite{Enke Liu}.  A large non-saturating MR is a characteristic feature of compensated semimetals.  At low fields the conventional MR ($B \parallel c$) is parabolic in nature and at higher fields it changes to a linear dependence. The parabolic MR in weak fields confirms the three dimensional nature of the charge carriers in Co$_{3}$Sn$_{2}$S$_{2}$.  The linear MR at high fields has been observed in most DSMs and WSMs but the origin of this behaviour is not yet understood. 

In order to confirm the chiral character in this semimetal we carried out the angle dependent MR at $T = 2$~K as shown in Fig.~\ref{fig:mr-hall}~(a) and longitudinal MR ($B \parallel I$) at various temperatures as shown in Fig.~\ref{fig:mr-hall}~(b).  It is clear from the data shown in Fig.~\ref{fig:mr-hall}~(a) and (b), that NLMR is absent in the $B \parallel I$ configuration at temperatures down to $T = 2$~K\@. The absence of NLMR in our sample might be due to the large value of the orbital MR in the $B\parallel c$ configuration.  It must be pointed out that the magnitude of the previously reported NLMR in $B \parallel I$ configuration was only $2\%$  and it was observed only at $T = 0.3$~K \cite{Enke Liu} which is almost an order of magnitude smaller than the lowest temperature of our measurements. Thus, the large ($160\%$) orbital MR observed for our Co$_3$Sn$_2$S$_2$ crystals, which is larger than previously reported \cite{Enke Liu}, can easily mask the relatively much smaller negative MR resulting from the chiral anomaly.  Since the Chiral anomaly critically depends on $B$ being in the direction of $I$, we have checked for the misalignment of current and magnetic field by rotating the sample up to $ 5^\circ $ in either direction of $B \parallel I$.  However, we did not observe a negative MR in our experiment at low temperature.  We do observe a negative MR at higher temperatures close to $T_c = 175$~K\@. We believe that this negative MR is due to suppression of spin disorder scattering in magnetic fields.

The Hall effect data taken at different temperatures is shown in the Fig.~\ref{fig:mr-hall}~(c) and (d). In order to remove the possible contribution of the orbital MR from the Hall data, the measured Hall data is anti-symmetrized. A sharp switching behaviour can be seen in the Hall resistivity at low temperatures, which is a signature of the anomalous Hall effect.  A rectangular hysteresis loop can be observed in Fig.~\ref{fig:mr-hall}~(d) which is the low field zoomed in version of the Hall data shown in Fig.~\ref{fig:mr-hall}~(c).  The coercive field at $2$~K is found to be $0.5$~T, which decreases with increasing temperature, as expected for a ferromagnetic material. The value of the anomalous Hall conductivity $\sigma_{H}^{A} $ is calculated by using the expression $\sigma_{H}^{A} = \rho_{H}^{A}/(\rho_{H}^{A})^{2} + \rho^{2})$, where $\rho_{H}^{A}$ and $\rho$ are the anomalous Hall resistivity at zero field and the zero field resistivity, respectively.  The value of the $\rho_{H}^{A}$ was calculated by extrapolating the linear high field part of the Hall resistivity to zero field.  At $2$~K the value of $\sigma_{H}^{A}$ is found to be {\color {blue} 1266}~$(\Omega~cm)^{-1}$.  Additionally, we observe nonlinearity in the Hall resistivity at high fields which indicates the existence of the two types of the carriers e.g electrons and holes. These results are in good agreement with recent reports on this compound \cite{Enke Liu,Qi Wang,Kassem3,Schnelle}.

We next present discovery of the PHE in Co$_{3}$Sn$_{2}$S$_{2}$.  The measurement geometry for the PHE consists of in plane current, Hall voltage leads, and magnetic field.  The magnetic field makes an in-plane angle $\phi$ with the current $I$.  Figure~\ref{fig:planar}~(a) and (c) show the planar Hall resistivity $\rho$$^{PHE}$ and longitudinal resistivity data as a function of $\phi$ at $T = 2$~K\@. In order to remove the contribution of the normal Hall resistivity from the $\rho$$^{PHE}$, we have averaged the $\rho_{xy}$ data measured at positive and negative magnetic fields.  It can be seen from the Fig.~\ref{fig:planar}~(a), that $\rho$$^{PHE}$ increases with increasing field.  As predicted theoretically for PHE, $\rho$$^{PHE}$ shows minima and maxima near $\phi = 45^\circ$ and $135^\circ$, respectively. The position of minima and maxima is consistent with previously reported PHE results on other DSMs such as Cd$_3$As$_2$, MoTe$_2$, GdPtBi, ZrTe$_5$,and WTe$_{2}$ \cite{P. Li, Li, Singha,Kumar,Chen F,Wu,Yang J,Najmul} .  The contribution from the chiral anomaly to $\rho$$^{PHE}$ and $\rho_{xx}$ is given by the expressions \cite{Nandy,Burkov1},

\begin{equation}
\centering \rho^{PHE} = -\Delta \rho^{chiral} \sin \phi \cos \phi + b 
\label{Eqn1}
\end{equation}

\begin{equation}\label{2}
\centering 
\rho_{xx} = \rho_{\perp}- \Delta \rho^{chiral} \cos^{2} \phi     
\end{equation}

where $\Delta \rho_{chiral} =  \rho _{\perp}-\rho_{\parallel}$ is the chiral anomaly induced resistivity and $\rho$$_{\perp}$ and $\rho$$_{\parallel}$ are the transverse ($\phi = 90^\circ$) and longitudinal ($\phi = 0^\circ$) resistivity in the planar Hall measurement geometry. The constant $b$ accounts for the anisotropic MR resulting from any small mis-alignment.  The data in Figs.~\ref{fig:planar}~(a) and (b) were fit by the above expressions and the chiral resistivity $\Delta \rho_{chiral}$ and the transverse resistivity $\rho_\perp$ were extracted at each magnetic field.  The fits to Eqn.~(1) and (2) are shown in the inset of Figs.~\ref{fig:planar}~(c) and (d), respectively.   The obtained $\Delta \rho_{chiral}$ and $\rho_\perp$ are plotted versus the magnetic field in Figs.~\ref{fig:planar}~(c) and (d).  Both $\Delta \rho_{chiral}$ and $\rho$$_{\perp}$ show a monotonically increasing trend with increasing field. We have fit the magnetic field dependence to a power law ($ B^{n} $). The extracted  value of the exponents for the two cases is found to be nearly quadratic ($n = 2$) for both $\Delta \rho_{chiral}$ and $\rho$$_{\perp}$.  These values are similar to exponents obtained for other DSM and WSM materials \cite{Singha}. 

\begin{figure*}
	\centering
	\includegraphics[width=0.85\linewidth]{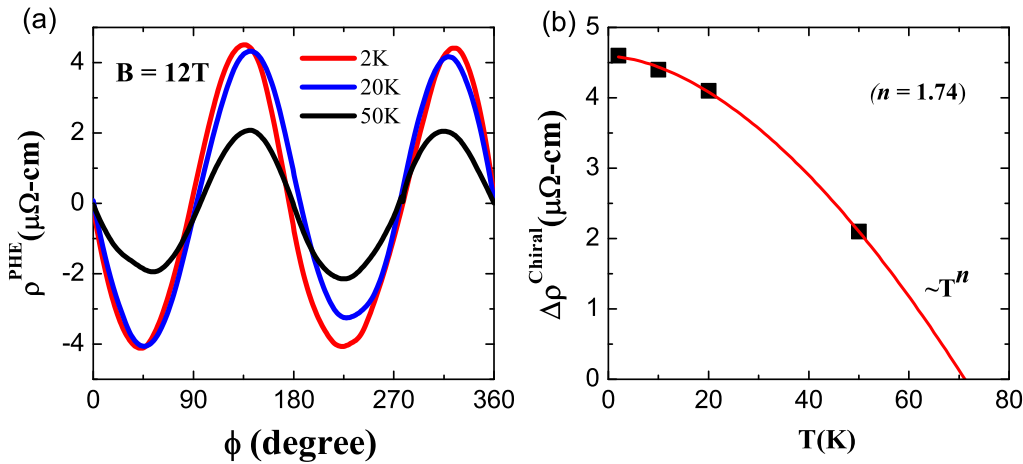}
	\caption{ (a) Planar Hall resistivity $\rho^{PHE}$ vs angle, measured at various temperatures in a magnetic field of $12$~T\@. (b) The extracted chiral contribution $\triangle$$\rho$$_{chiral}$ to the PHE vs temperature $T$.  The solid curve through the data is a power law fit.}
	\label{fig:chiral-T}
\end{figure*}

The data in Fig.~\ref{fig:planar} clearly demonstrates the presence of the PHE in Co$_3$Sn$_2$S$_2$.   However, given that the material shows a ferromagnetic ordering below $T_c = 175$~K, it is unclear whether the PHE is related to the topological character of the material, or to the ferromagnetism.  Previous studies of the PHE in ferromagnetic materials have shown that the PHE if present due to ferromagnetism, tracks the magnetization and is observed in the ordered state but vanishes at the ferromagnetic critical temperature [Fig.3c in Ref.\onlinecite{Tang2003} for example].  We have therefore tracked the temperature dependence of the PHE in Co$_3$Sn$_2$S$_2$.  Figure~\ref{fig:chiral-T}(a) show the temperature dependent $\rho^{\rm PHE}$ measured in $12$~T at various temperatures and Fig.~\ref{fig:chiral-T}(b) shows the chiral contribution extracted from these data.  It can clearly be seen that the chiral contribution decreases on increasing the temperature.  We performed a phenomenological fit to the data by the expression $\rho^{\rm chiral} = A+BT^n$, where $A, B$, and $n$ are fit parameters.  This fit, shown as the solid curve through the data in Fig.~\ref{fig:chiral-T}(b) extrapolates to $\rho^{\rm chiral} = 0$ at $T \approx 71$~K\@.  This clearly demonstrates that the PHE signal vanishes at a temperature much smaller than the ferromagnetic ordering temperature $T_c = 175$~K, thereby suggesting that the origin of the PHE is not related to the ferromagnetism.

\section{Summary}
In conclusion, we have demonstrated the Topological character in the recently discovered magnetic Weyl semimetal Co$_{3}$Sn$_{2}$S$_{2}$ by observation of the planar Hall effect (PHE) starting at temperatures ($T \leq 70$~K) much below the ferromagnetic ordering temperature $T_c = 175$~K\@.  Along with the large intrinsic anomalous Hall conductivity of the order of $1266~(\Omega~cm)^{-1}$ found in our measurements, this provides strong evidence for the topological band structure of this compound. It must be noted that the PHE is observed even though we did not observe the chiral anomaly induced NLMR.  We point out that the NLMR reported previously was observed at very low temperature ($T = 0.35$~K) and is only $2\%$ in magnitude \cite{Enke Liu} which is quite weak and could be easily masked by a large orbital MR as was observed in our measurements.  It has recently been pointed out theoretically that the planar Hall effect is a more direct evidence of the nontrivial band structure of a Weyl semimetal \cite{Nandy,Burkov1}.  Our results indeed demonstrate that the PHE can be used to study the Topological character of a material even if extrinsic effects mask the Chiral anomaly induced negative MR.\\

\noindent
\emph{ Acknowledgments.--} We acknowledge use of the x-ray and SEM facilities at IISER Mohali. We thank Dr. Suvankar Chakraverty from INST, Mohali for the use of his PPMS for some of our measurements.

\end{document}